

\magnification\magstep 1

\line{\hfil NSF-ITP-93-35}

\vskip 0.4truecm
\centerline{\bf Effective description of axion defects}
\vskip 1truecm
\centerline{P.K. Townsend}
\vskip 1 cm
\centerline{ITP, Univ. of California at Santa Barbara,}
\centerline{California, U.S.A.}
\vskip 0.8truecm
\centerline{\it and}
\vskip 0.8truecm
\centerline{DAMTP, Univ. of Cambridge,}
\centerline{Silver St., Cambridge, U.K.}
\vskip 1cm

\centerline{\bf ABSTRACT}
\vskip 0.5cm
Axion strings and domain walls exhibit a number of
novel effects in the presence of
gauge fields, in particular the electromagnetic field.
It is shown how these
effects are reproduced in a model
of Nambu-Goto-type strings and open or
closed membranes coupled to gauge fields.
The generalization to `axionic
p-branes' is considered and it is shown
how worldvolume gauge fields that arise in
certain cases can be interpreted as Goldstone fields.

\vfill\eject

Pseudoscalar fields that couple to a
topological charge density are generically
known as axion fields. For most of this paper we
consider an axion field
$\theta(x)$ coupling to the electromagnetic two-form
field strength $F=dA$ via an interaction of the form
$$
{\cal L} = {n e^2\over 8\pi^2} \int_{{}_{\cal M}}\theta\,
F\wedge F
\eqno (1)
$$
where ${\cal M}$ is a four-dimensional spacetime, $e$ is the
electromagnetic coupling constant and $n$ is the axion
coupling constant which, it will turn out,
must be an integer. A further requirement on $\theta$ is that
it take values in $S^1$, which is achieved by making the
identification $\theta\sim \theta +2\pi$.
Given this identification, configurations are
possible for which $\theta(x)$ changes by $2\pi$ upon traversal
of a closed loop in space. Such a loop is
threaded by a one-dimensional topological defect known as
an axion string. The physics of axion strings depends
on whether the axion is massless or massive.
If it is massless then the string configuration will attempt
to minimize the gradient energy density
$(\nabla\theta)^2$, so $\theta$ will
change smoothly as the string is circumnavigated.
On the other hand, if there is an axion potential with a
single global minimum in the interval $(\theta,\theta+2\pi)$,
at $\theta=\theta_0$ say, then $\theta$ will prefer to
remain at $\theta_0$ everywhere except in a small region,
through which it changes by $2\pi$. This region is
the core of an axion domain wall interpolating between
$\theta_0$ and $\theta_0+2\pi$, and the axion string is
the boundary of this wall.

Callan and Harvey have shown that one consequence of
the interaction term (1) for axion strings is
that an electric field with a non-zero component
parallel to the string causes a net inflow of electric charge
onto the string [1]. Axion domain walls are also
associated with novel electromagnetic effects. For
example, Sikivie [2], and subsequently Lee [4] and
Wilczek [5], have shown that, owing to the
$\theta$-dependence of the electric charge of a magnetic
monopole [3], a monopole that passes adiabatically through
a domain wall is converted into a dyon. Thus there is a charge
exchange between monopoles or dyons and axion domain walls.
This effect can be understood in terms of an electric
charge that is induced on an axion domain
wall by a transverse magnetic field.
Another consequence of this induced charge
is that a spherical domain wall enclosing a monopole
carries a unit electric charge which can stabilize it [6].
This is reminiscent of Dirac's membrane model of
the electron [7] in which, however, the electric charge was
added in an {\sl ad hoc} manner, there being no
monopole in the interior to induce it.  The Callan-Harvey
(CH) effect and the Sikivie effect are closely related
since the latter can be used to derive the
former in the case that the string is the boundary of
a domain wall. To see this observe that one way of
producing an electric field along the string is to cycle a
magnetic monopole around it. Each time the monopole
passes through the wall it transfers one unit of charge
to the wall and the net charge transfer is that
predicted by the CH effect [8].

The principal purpose of this paper is to explain how
these effects can be understood in terms of an
effective theory of Nambu-Goto strings and Dirac
membranes coupled to the electromagnetic field.
There are three cases to consider: (i) a closed membrane,
representing a domain wall without a boundary,
(ii) an open membrane, for which the boundary represents
an axion string, and (iii) a closed string
which is {\it not} the boundary of a domain wall,
representing a string defect in a massless axion field.
This work was begun with the hope that the subsequent
extension to higher-dimensional objects in a
higher-dimensional spacetime might help in the
construction of the worldvolume action for the type
II p-branes that have recently been found as solutions of
$d=10$ $N=2$ supergravity theories [9,10,11], and
was motivated in part by the observation of Witten
that the heterotic string can be viewed as an axion string
[12]. In the event, the chief insight gained is an
interpretation of the worldvolume gauge fields that
appear in these models as Goldstone fields, and the
consequent determination of certain of their couplings
to spacetime fields, as explained at the conclusion of this work.

We begin with case (i), some aspects of which were
discussed in [4]. The worldvolume fields $X^\mu(\zeta)$
define a map $\phi: W\rightarrow {\cal M}$ from
the three-dimensional worldvolume $W$ swept out
by a closed membrane $M$, with coordinates $\{\zeta^I\}$.
In the absence of the electromagnetic field we
expect the effective action to be that of
Dirac's relativistic membrane,
i.e. the volume of $W$ in the metric induced on it
by the spacetime metric. To
obtain the required electromagnetic interaction of this
membrane, observe that the
field theory interaction Lagrangian density implicit
in (1) can be written as
$$  {\cal L} = -{ne^2\over 8\pi^2}\partial_\mu \theta
\varepsilon^{\mu\nu\rho\sigma} A_\nu\partial_\rho A_\sigma \ .
\eqno (2)
$$
Choose local coordinates such that $\theta=\theta(z)$ across the
wall. Then performing the $z$-integration we conclude,
following [4], that the action governing the
electromagnetic interactions of the membrane is
$$
S_{int} ={ne^2\over 4\pi}\int_{{}_W}\!\! d^3\zeta\,
\varepsilon^{IJK}A_I\partial_J
A_K  \eqno (3)
$$
where $A_I\equiv \partial_I X^\mu A_\mu$.
For a {\sl closed} membrane this
action is invariant under the electromagnetic gauge transformation
$$
A_\mu\rightarrow A_\mu +\partial_\mu \Lambda\ ,
\eqno (4)
$$
as required. From (3) we deduce that the
electric current on the wall is
$$
J_{{}_M}^\mu(x) \equiv {\delta S_{int}\over\delta A_\mu(x)}
={ne^2\over
2\pi}\int_{{}_W}\!\! d^3\zeta\; \varepsilon^{IJK}
\partial_J A_K \partial_IX^\mu\,
\delta^{(4)}\big(x-X(\zeta)\big)\ .
\eqno (5)
$$
This current is identically conserved (except at
the initial and final times, where the membrane
may be considered as being created and destroyed).
The total charge on the membrane is therefore
$$
Q\equiv \int\! d^3 x J_{{}_M}^0(x) =
{ne^2\over 2\pi}\int_{{}_W}\!\!d^3\zeta\,
\varepsilon^{IJK}\partial_IX^0\partial_J A_K\,
\delta\big(t-X^0(\zeta)\big)\ .
\eqno (6)
$$
Performing the $\zeta^0$-integral (e.g. in the
$X^0=\zeta^0$ gauge) we find, in
the presence of a magnetic field ${\bf B}$, that $Q$ is
given by the surface integral
$$
Q= {ne^2\over 2\pi}\oint\!{\bf dS}\cdot {\bf B} =
ne \big( {eg\over
2\pi}\big) \ ,
\eqno (7)
$$
where $g$ is the enclosed magnetic charge. According to
the Dirac quantization condition the minimal
magnetic charge satisfies $eg=2\pi\hbar$,
in which case $Q= ne\hbar$. Consistency with
charge quantization now requires that the axion
coupling constant $n$ be an integer.
For $n=1$ we see that a monopole induces
one unit of electric charge on an axion
domain wall that encloses it, in agreement
with [6]. Note, however, that if
$\theta$ is identified with $\theta +2\pi$ the wall
of such a `monopole bag' can
be expected to be unstable against the nucleation
of holes bounded by axion strings.

An alternative way to determine the charge
on a closed membrane is to determine
the transformation of its wavefunctional
$\Psi$ under a time-dependent gauge
transformation $A_{{}_0} \rightarrow A_{{}_0}+
\partial_{{}_0}\Lambda(\tau)$. Let
$\Psi$ at time $\tau_f$ be expressed as a
path integral of ${\rm
exp}\big({(i/\hbar)S}\big)$ over a three-manifold
with two boundaries representing
the membrane at time $\tau_f$ and an earlier time $\tau_i$.
The interaction term
(3) in the action $S$ is {\sl not} invariant under
a time dependent gauge
transformation, but rather transforms as
$$
S_{int} \rightarrow S_{int} + {ne^2\over 2\pi}[\Lambda]
\oint {\bf dS}\cdot {\bf B}
\eqno (8)
$$
where $[\Lambda]$ is the difference in $\Lambda$ between
the initial and
final times. (Note the factor here! The procedure of
taking $A_I\rightarrow A_I
+\partial_I\Lambda$ in (3) and then specializing to
$\Lambda=\Lambda(\tau)$,
which yields a result differing by a factor of $2$,
is incorrect if $g\ne0$
because  ${\bf A}$ is then not globally defined). Using
again the Dirac quantization condition
$eg=2\pi\hbar$, we conclude that
$$
\Psi\rightarrow e^{ine[ \Lambda]}\Psi \ ,
\eqno (9)
$$
exactly as for a particle
with charge $n$ times the electric charge. Given
that the electromagnetic group is
$U(1)$, we should make the identification
$\Lambda\sim \Lambda +{2\pi\over e}$, in
which case we again see that $n$ must be an integer.

When a monopole inside a closed domain wall
passes passes through the wall the
induced charge must be transferred to the monopole.
To see how this happens
in the present context, observe that applied electric
and magnetic fields induce
the surface electric charge and current densities
$$
\rho = {ne^2\over 2\pi}{\bf n}\cdot {\bf B}\qquad {\bf j}
= {ne^2\over 2\pi} {\bf
n}\times {\bf E}
\eqno (10)
$$
where ${\bf n}$ is normal to the surface. Maxwell's
equation $\nabla\times{\bf E}
-\dot{\bf B} ={\bf j}_{mag}$, for magnetic current
density ${\bf j}_{mag}$,
implies that
$$
\nabla\cdot {\bf j} +\dot\rho = -{ne^2\over 2\pi}{\bf n}\cdot
{\bf j}_{mag}
\eqno (11)
$$
i.e. charge is locally conserved on the
membrane {\sl except when pierced by a
magnetic charge}. Integrating over a closed membrane
and using the continuity
equation for magnetic charge we find that the rate of
change of total charge $Q$
on the membrane is given by $\dot Q= {ne^2\over 2\pi}\dot g$.
Hence $\Delta
Q=ne\hbar$ is the change in charge on the membrane
when a monopole passes through
it.

There is a striking similarity here with a
result of Lee for $d=3$
Chern-Simons/Higgs theories [13]. He showed that instantons ('t
Hooft-Polyakov monopoles from the d=4 perspective)
violate charge conservation
by $k$ units, $k$ being the integer multiple of
the Chern-Simons density. A
monopole passing through a domain wall could therefore
be interpreted as an
instanton by a `flatlander' living in the wall.

We turn now to case (ii), the open membrane. The
electric current (5) now
acquires an additional contribution because of the
membrane boundary but is
nevertheless not conserved at the boundary,
indicating that the
electromagnetic gauge invariance is broken there.
However, since the
underlying axion field theory {\sl is} gauge-invariant
this breakdown of gauge
invariance must be spontaneous. The effective theory
will therefore include an
additional variable $y$, taking values in $S^1$ and
defined on the worldsheet $w$
swept out by the boundary of $M$ (i.e. the timelike
component of $\partial W$)
with local coordinates $\{\xi^i\}$, and with an
action that restores
electromagnetic gauge invariance when the
transformation $y\rightarrow y+\Lambda$
is taken into account. It is readily verified that
the following action satisfies
this requirement:  $$  S= S_{{}_M} + {ne^2\over
4\pi}\int_w\! d^2\xi\,\big\{ \varepsilon^{ij}
\partial_i yA_j - \sqrt{-g}[
{\rm const.}\,  + {1\over2}g^{ij} D_iyD_jy]\big\} \ ,
\eqno (12)
$$
where $S_{{}_M}$ is the volume part of the membrane
action, which includes
(3), $g^{ij}$ is the inverse of the induced
metric $g_{ij}$ and $g=\det g_{ij}$,
and
$$
D_i y\equiv \partial_i y - A_i
\eqno (13)
$$
(where $A_i\equiv \partial_iX^\mu A_\mu$)
is the covariant derivative of the Goldstone
variable $y$. The term in
(12) proportional to the constant is necessary
because otherwise every point of
$\partial M$ would move at the speed of light.
The electromagnetic current
is now given by
$$
J^\mu(x) = J_{{}_M}^\mu + J_{{}_{\partial M}}^\mu\ ,
\eqno (14)
$$
where $J_{{}_M}^\mu$ is given in (5) and
$$
J_{{}_{\partial M}}= {ne^2\over 4\pi}\int_w\!\! d^2\xi\,\big[
\varepsilon^{ij}D_i y + \sqrt{-g}g^{ij}D_iy\big]\partial_j
X^\mu  \delta^{(4)}\big( x-X(\xi)\big)\ .
\eqno (15)
$$
Note that this surface contribution to the current
has contributions both from the
new surface term in the action {\sl and} from the
volume term (as a result of an
integration by parts). Using the $y$ equation of
motion that follows from (12) one
finds that $\partial_\mu J^\mu=0$, as expected from
gauge invariance.
If now an electric field parallel to a domain wall
boundary is applied
then equations (10) imply that charge will
accumulate on the boundary. This is  the
CH effect in the case that the axion string
is the boundary of a domain wall.

We now consider case (iii), a closed string. One would
expect the action to
include the boundary terms of the open
membrane action, but this cannot be the
complete action because the boundary current (15)
is not conserved. There must be
a correction to this current from the axion
field from which the string was
formed, but which can no longer be localized
on a domain wall. If one recalls
that a massless axion field is equivalent to
a second-rank antisymmetric gauge
potential $B_{\mu\nu}$ (for a four-dimensional spacetime)
and that such a field has
a natural coupling to a string, then it is clear that the
solution to the problem must involve this
coupling. We are therefore led to
consider the closed string action
$$
S= -{ne^2\over 8\pi}\int_w\!\! d^2\xi\bigg\{ \sqrt{-\gamma}
\gamma^{ij}\big( g_{ij}
+  D_iyD_jy\big) - \varepsilon^{ij}\big(
B_{ij} + 2\partial_i y A_j\big)\bigg\}
\ ,
\eqno (16)
$$
where $B_{ij}$ is the pullback to the
worldsheet of $B_{\mu\nu}$, and we have
introduced a new independent worldsheet
metric $\gamma_{ij}$. [The
Euler-Lagrange equation for $\gamma_{ij}$
implies that it is conformally
equivalent to $g_{ij} + D_iyD_jy$, so elimination
of $\gamma$ in favour of the
induced metric $g$ will introduce quartic and higher
terms in $Dy$ that we did not
previously include (although we could have done
so since they are separately
gauge-invariant); but higher derivative terms are
irrelevant to a low energy
effective action and so we need not distinguish
between actions which differ by
them]. The action (16) is gauge-invariant provided
$B_{\mu\nu}$ transforms as
$$
B_{\mu\nu}\rightarrow B_{\mu\nu} +
2A_{[\mu}\partial_{\nu]}\Lambda
\eqno (17)
$$
under an electromagnetic gauge transformation. Note that this
transformation follows [14] from the identification
of $B_{\mu\nu}$ as the
dual of $\theta$ and the axion coupling (1), so there
is actually no need to
postulate it.

The `anomalous' transformation of $B_{\mu\nu}$ means that we
should now redefine the electric current to be
$$
\eqalign{
J^\mu(x) &\equiv {\delta S\over\delta A_\mu(x)} -
2A_\nu(x){\delta S\over\delta B_{\mu\nu}(x)}\cr
&= J^\mu_{{}_{\partial M}} +{ne^2\over 2\pi} \int d^2\xi\,
\varepsilon^{ij}A_i\partial_jX^\mu \,
\delta^{(4)}\big(x-X(\xi)\big)\cr}
\eqno (18)
$$
where $J^\mu_{{}_{\partial M}}$ is the current
given in (15). The new
contribution, which replaces $J_{{}_M}^\mu$,
ensures that the total current
is conserved.

We now discuss the non-abelian generalization
of these results. We
replace (1) by
$$
{\cal L} = {k \over 8\pi^2}
\int_{{}_{\cal M}}{\rm tr}(F\wedge F)\ ,
\eqno (19)
$$
where $F=dA +A^2$ is a Yang-Mills (YM) field
strength taking values in the
algebra $\bar G$ of a semi-simple group $G$.
For simplicity, we assume that
$A= A^aT_a$ with ${\rm tr}(T_aT_b)=(1/2)\delta_{ab}$.
The indentification
$\theta\sim \theta +2\pi$ now forces the coupling
constant $k$ to be an integer.
This is analogous to the quantization condition
that we found for $n$ in the
abelian case, but the reasoning is quite
different since the quantization of $n$
did not depend on the indentification of $\theta$
with $\theta +2\pi$. This is
because axion domain walls (although not axion strings)
are possible even if this
identification is {\sl not} made; all one needs
is a potential $V(\theta)$
periodic in $\theta$.

Following the same reasoning as before, we deduce that
the membrane representing
the axion domain wall must couple to the gauge fields
via the {\sl induced}
Chern-Simons (CS) term [4]
$$
S_{CS} = {k\over 4\pi} \int_W\!
\phi^*{\rm tr}(AdA +{2\over3} A^3)\ .
\eqno (20)
$$
where $\phi^*$ is the pullback of the map $\phi$.
The quantization condition on $k$ is now recoverable
from the requirement that
(20) be invariant (for a closed membrane)
under gauge-transformations not
connected to the identity. That is, if $k$ fails
to be an integer we have
a global `sigma-model anomaly' of the same type
that enforces the quantization of
the coupling constant of the usual CS action [15],
but with a different
interpretation here since the gauge fields
are not defined as independent
worldvolume fields.

If the membrane has a boundary then the
action (19) will again
fail to be gauge-invariant and we must introduce
a boundary action depending on a
Goldstone variable $y$ taking values in the group $G$.
This boundary action will
consist of kinetic terms, as before,
which are manifestly gauge-invariant,
together with the additional term
$$
S_{{}_{GWZ}} = {k\over 4\pi}\int_w\!
\phi^*\big({\rm tr}(LA) +b\big)
\eqno (21)
$$
where $L$ is the left-invariant one-form on $G$
taking values in $\bar G$, and
$b$ is a two-form potential on G such that,
locally, $db={\rm tr} L^3$, i.e.
$S_{{}_{GWZ}}$ is a (`sigma-model' version of) a
Gauged-Wess-Zumino term for $G$.
That this is the appropriate term follows from
the fact that its YM variation
with $\bar G$-valued parameter $\epsilon$,
$$
\delta S_{{}_{GWZ}} = -{k\over 4\pi} \int_w \phi^*{\rm
tr}(Ad\epsilon) \ ,
\eqno (22)
$$
precisely cancels the variation of the volume-term
of the open membrane
action. The combined action is therefore
invariant. The same result was found
previously by Stone in his study of
edge states of a two-dimensional
droplet of fermi liquid in the quantum-Hall regime [16].

If the axion is massless and the axion string
therefore not the boundary of a
domain wall then, as in the abelian case, the
membrane action must be replaced by
the coupling to the string of the two-form gauge
potential $B$, dual to the axion
field and having an anomalous YM transformation
proportional to ${\rm
tr}(Ad\epsilon)$. At this point the axion string
action so obtained may be
identified as that appearing (for a d-dimensional
spacetime and gauge group
$E_8\times E_8$) in the group manifold formulation
of the heterotic string [17].
In effect, what has been shown here is that there
are two ways to cancel
sigma-model anomalies in string theory.
One can either couple the string to the
dual of an axion field {\sl or} one can regard
it as the boundary of a domain
wall.

Finally, we turn to higher-dimensional
analogues of axion strings. Defects in a
massless axion field have a straightforward
generalization to $d$-dimensional
spacetime and more general topological
charge densities, e.g.
${\rm tr}\big(F^{n+2}\big)$ for $n\ge 0$. In this case
the massless axion field
$\theta$ is replaced by a $(d-2n-4)$-form potential
$B$ with a coupling of the form
$$
\int_{{\cal M}_d} \!\! B\wedge {\rm tr}\big(F^{n+2}\big)\ .
\eqno (23)
$$
Since $B$ is defined only locally, the integral
of its field strength $H=dB$ over
a $(d-2n-3)$-sphere in $(d-1)$-space may not vanish.
Such an $S^{d-2n-3}$ must
then encircle a defect of dimension $(2n+1)$.
The gauge-invariant effective
action for this defect will then include (i)
a coupling of the induced YM
potential to a Goldstone variable taking values
in the YM group, and (ii) a
coupling to the dual $(2n+2)$ form potential $\tilde B$.
An action for a
$(2n+1)$-brane coupled to spacetime YM fields and
having these properties was
recently constructed [18]. Here we see that this
action can be interpreted as the
effective action of an axionic defect.

An alternative generalization involves the
replacement of the YM or
electromagnetic field by an $(n+1)$-form potential
$C$ with field strength $F=dC$
and a coupling to the axionic $(d-2n-4)$-form
potential $B$ via the term
$$
\int_{{\cal M}_d} \!\! B\wedge F\wedge F\ .
\eqno (24)
$$
In this case the worldvolume action for the axion
defect of dimension $(2n+1)$
will include a term of the form
$$
\int_{W_{2n+2}}\!\!\phi^*\big( \tilde B +
2dY\wedge C\big)\ ,
\eqno (25)
$$
where $Y$ is an $n$-form Goldstone variable with
the transformation $\delta Y=
\Omega$ under the gauge transformation
$\delta C=d\Omega$ of $C$, and $\tilde B$
has the `anomalous' transformation
$\delta\tilde B=-2d\Omega\wedge C$.
The special case $d=10$ and $n=2$ is of particular
interest because the $d=10$
$N=2a$ supergravity theory has a coupling of
the form (24) and an axionic
fivebrane defect [9]. We conclude from the above
analysis that its
six-dimensional worldvolume action must
involve a two form potential.
This is known to be the case [9]. We now see that
{\sl all} of the worldvolume
fields of the type IIa fivebrane have an
interpretation as Goldstone fields. We
have also determined some features of how the
worldvolume two-form potential
couples to spacetime fields, but the story is far
from complete since, among other
reasons, the supergravity three-form $C$ has its
own `anomalous' transformation
under an additional $U(1)$ gauge group and the
action (25) therefore fails to be
$U(1)$ invariant.

The $d=10$ $N=2b$ supergravity has a four-form
potential $A_{{}_4}$ with a
self-dual five-form field-strength
$F_{{}_5}=dA_{{}_4}$ and {\sl two} two-form
potentials $B_{{}_2}^{(a)}$ (a=1,2) with
three-form field-strengths
$H_{{}_3}^{(a)}$. Because of the self-duality
of $F_{{}_5}$ no Lorentz-invariant
action is known but one can deduce from the
field-equations [19] that (in one of
its possible dual formulations) it should contain the term
$$
\int_{{\cal M}_{10}}\!\! B_{{}_2}^{(1)}
\wedge H_{{}_3}^{(2)}\wedge F_5 \ .
\eqno (26)
$$
Axionic defects are possible with either
$B_{{}_2}^{(2)}$ (say) or $A_{{}_4}$
taking the role of the axion field [9-11]. This leads,
respectively, to the type
IIb fivebrane and threebrane. In either case the
effective worldvolume action will
require a Goldstone one-form potential $Y_1$,
in agreement with [9-11]. We
now see that this worldvolume gauge potential
must transform as
$\delta Y_1=\Omega_{{}_1}$ under the spacetime
gauge transformation $\delta
B_{{}_2}^{(1)}= d\Omega_{{}_1}$. For the threebrane,
for example, this implies
that the coupling to $N=2b$ supergravity will
require a term of the form
$$
\int_{W_4}\! \phi^* \big(A_4 + 2dY_1\wedge B_2^{(2)}\big)
\eqno (27)
$$
in the four-dimensional worldvolume action.

\vskip 0.5cm
\leftline{\bf Acknowledgements}
\vskip 0.3cm
I am grateful to Michael Stone and Andy Strominger
for discussions and to the
members of the ITP at UCSB for their hospitality.
This work was supported in part
by the National Science Foundation grant no. PHY89-04035.

\vskip 1cm
\centerline{\bf References}
\vskip 1cm
\item {[1]}
C.G. Callan and J.A. Harvey, Nucl. Phys. {\bf B250} (1985) 427.

\item {[2]}
P. Sikivie, Phys. Lett. {\bf 137B} (1984) 353.

\item {[3]}
E. Witten, Phys. Lett. {\bf 86B} (1979) 283.

\item {[4]}
K. Lee, Phys. Rev. {\bf D35} (1987) 3286.

\item {[5]}
F. Wilczek, Phys. Rev. Lett. {\bf 58} (1987) 1799.

\item {[6]}
I.I. Kogan, Phys. Lett. {\bf B299} (1993) 16.

\item {[7]}
P.A.M. Dirac, Proc. Roy. Soc. London Ser.{\bf A268} (1962) 57.

\item {[8]}
S.G. Naculich, Nucl. Phys. {\bf B296} (1988) 837.

\item {[9]}
C. Callan, J. Harvey and A. Strominger, Nucl. Phys.
{\bf B359} (1981) 611; Nucl. Phys. {\bf B367} (1991) 60.

\item {[10]}
G.T. Horowitz and A. Strominger, Nucl. Phys. {\bf B360} (1991) 197.

\item {[11]}
M.J. Duff and J.X. Lu, Phys. Lett. {\bf B273} (1991) 409; Nucl.
Phys. {\bf B390} (1993) 276.

\item {[12]}
E. Witten, Phys. Lett. {\bf 153B} (1985) 243.

\item {[13]}
K. Lee, Nucl. Phys. {\bf B373} (1992) 735.

\item {[14]}
H. Nicolai and P.K. Townsend, Phys. Lett. {\bf 98B} (1981) 257.

\item {[15]}
S. Deser, R. Jackiw and S. Templeton, Phys. Rev. Lett.
{\bf 48} (1982) 975; Ann. Phys. (N.Y.) {\bf 140} (1982) 372.

\item {[16]}
M. Stone, Ann. Phys. (N.Y.) {\bf 207} (1991) 38.

\item {[17]}
M.J. Duff, B.E.W. Nilsson and C.N. Pope, Phys.
Lett. {\bf 163B} (1985)
343;$\qquad$ R.~Nepomechie, Phys. Lett.
{\bf 171B} (1986) 195; Phys. Rev. {\bf
D33} (1986) 3670; M.J. Duff, B.E.W. Nilsson,
C.N. Pope and N.P. Warner, Phys.
Lett. {\bf 171B} (1986) 170; ; R. Kallosh,
Phys. Lett. {\bf 176B} (1986) 50.

\item {[18]}
J.A. Dixon, M.J. Duff and E. Sezgin, Phys. Lett.
{\bf 279B} (1992) 265.

\item {[19]}
J.H. Schwarz, Nucl. Phys. {\bf B226} (1983) 269.

\end